\documentclass[12pt, a4paper]{article}
\usepackage{epsf}
\usepackage{cite}
\usepackage{amsmath,amssymb}
\input{colordvi.tex}
\usepackage[usenames,dvipsnames]{color}
\usepackage{graphicx}
\usepackage{ascmac}
\usepackage{hyperref}

\begin{document}

\begin{titlepage}

\begin{center}
\hfill TU-1181\\
\hfill KEK-QUP-2023-0004
\vskip 1.0in

{\Large \bf
Viable Vector Coherent Oscillation Dark Matter
}

\vskip .5in

{\large
Naoya Kitajima$^{(a,b)}$ and
Kazunori Nakayama$^{(b,c)}$
}

\vskip 0.5in

$^{(a)}${\em 
Frontier Research Institute for Interdisciplinary Sciences, Tohoku University, Sendai 980-8578, Japan
}

\vskip 0.2in

$^{(b)}${\em 
Department of Physics, Tohoku University, Sendai 980-8578, Japan
}

\vskip 0.2in

$^{(c)}${\em 
International Center for Quantum-field Measurement Systems for Studies of the Universe and Particles (QUP), KEK, 1-1 Oho, Tsukuba, Ibaraki 305-0801, Japan
}

\end{center}
\vskip .5in

\begin{abstract}

We construct a viable model of the vector coherent oscillation dark matter. The vector boson is coupled to the inflaton through the kinetic function so that the effective Hubble mass term is cancelled out. In order to avoid strong constraints from isocurvature perturbation and statistically anisotropic curvature perturbation, the inflaton is arranged so that it does not contribute to the observed large scale curvature perturbation and we introduce a curvaton. We found viable vector coherent oscillation dark matter scenario for the wide vector mass range from $10^{-21}\,{\rm eV}$ to $1\,{\rm eV}$.

\end{abstract}

\end{titlepage}

\tableofcontents

\section{Introduction}

Dark photon (or simply we may call it as a vector boson) is one of the plausible candidates of dark matter (DM) and many experiments and observations are searching for the evidence of dark photon DM~\cite{Fabbrichesi:2020wbt,Caputo:2021eaa}. 
There are various known production mechanisms of dark photon DM:
gravitational production~\cite{Graham:2015rva,Ema:2019yrd,Ahmed:2020fhc,Kolb:2020fwh,Sato:2022jya,Redi:2022zkt}, gravitational thermal scattering~\cite{Tang:2017hvq,Garny:2017kha}, misalignment production that we will discuss below~\cite{Nelson:2011sf,Arias:2012az,Nakayama:2019rhg,Nakayama:2020rka}, production through axion-like couplings~\cite{Agrawal:2018vin,Co:2018lka,Bastero-Gil:2018uel}, Higgs dynamics~\cite{Dror:2018pdh,Nakayama:2021avl}, kinetic couplings~\cite{Salehian:2020asa,Firouzjahi:2020whk,Nakai:2022dni},\footnote{The scenario of Ref.~\cite{Nakai:2020cfw} is not consistent with the constraint from the DM isocurvature perturbation.} and cosmic strings~\cite{Long:2019lwl,Kitajima:2022lre}. 
These production mechanisms do not require the kinetic mixing between the dark photon and the Standard Model photon. The production through the kinetic mixing from the thermal plasma has been discussed in Refs.~\cite{Jaeckel:2008fi,Pospelov:2008jk,Redondo:2008ec,An:2013yfc,Redondo:2013lna}, with a conclusion that this scenario cannot account for the observed amount of DM without violating observational bounds on the kinetic mixing parameter.\footnote{
	See Ref.~\cite{Gan:2023wnp} for the case of varying kinetic mixing parameter.
}
In this paper we revisit the misalignment production mechanism of dark photon DM.

Let us summarize a situation of misalignment production of dark photon DM.
The first proposal of Ref.~\cite{Nelson:2011sf} in a minimal massive vector boson model does not work actually, since the vector boson condensate exponentially damps during inflation due to the effective Hubble mass term in a physical basis.
The introduction of $(1/12)R A_\mu A^\mu$ term with $R$ being the Ricci scalar to cancel such a Hubble mass term, as proposed in Ref.~\cite{Arias:2012az}, also suffers from serious ghost instability problem, as pointed out in Ref.~\cite{Nakayama:2019rhg}.\footnote{
	The existing of ghost instability in some nonminimal vector boson theories had already been pointed out in Refs.~\cite{Himmetoglu:2008zp,Himmetoglu:2009qi} in the context of magnetogenesis~\cite{Turner:1987bw} and vector curvaton models~\cite{Dimopoulos:2006ms}.
}
Ref.~\cite{Nakayama:2019rhg} proposed a kinetic function model in which the inflaton $\phi$ and vector boson couples like $f^2(\phi) F_{\mu\nu} F^{\mu\nu}$ in order to cancel the effective Hubble mass term, where $f(\phi)$ is called the kinetic function, along the line of vector curvaton models~\cite{Dimopoulos:2007zb,Dimopoulos:2009am,Dimopoulos:2009vu,Wagstaff:2010qhd}.
However, later the same author pointed out that the vector coherent oscillation DM scenario is observationally excluded by the combination of constraints from the magnitude of DM isocurvature perturbation and the statistical anisotropy of the curvature perturbation~\cite{Nakayama:2020rka}. 
These two constraints are complementary and no viable parameter regions are found. 
The reason for this complementarity is that one needs small inflationary scale to suppress the DM isocurvature perturbation, as is well known in the context of axion DM, while small inflationary scale requires a large inflaton-vector coupling through the kinetic function $f(\phi)$ in order to cancel the effective Hubble mass term, which leads to large statistical anisotropy in the power spectrum of the curvature perturbation.

At the same time, Ref.~\cite{Nakayama:2020rka} also briefly pointed out several loopholes. One of the loopholes is to liberate $\phi$ from taking a role of the origin of the curvature perturbation. Then the constraint from the statistical anisotropy is significantly relaxed. 
The observed large scale curvature perturbation is generated by another scalar field, called the curvaton~\cite{Enqvist:2001zp,Lyth:2001nq,Moroi:2001ct}.
In this paper we will revisit this scenario in more detail.
We show that there are still nontrivial constraints from the statistical anisotropy and isocurvature perturbation through the measurement of cosmic microwave background (CMB) anisotropy, but we find viable parameter spaces.

This paper is organized as follows. In Sec.~\ref{sec:vec} we present the basic dynamics of the vector boson and inflaton during and after inflation in our scenario. The vector boson abundance is calculated. In Sec.~\ref{sec:const} various cosmological constraints on our model will be explained. Our scenario predicts many unusual properties of the cosmological perturbations, including statistical anisotropy in the curvature/isocurvature perturbation due to the presence of inflaton and curvaton. The isocurvature perturbation also has both an intrinsic and residual components, which need to be treated carefully. 
Combining all these constraints, we will find viable parameter ranges for successful vector coherent oscillation DM scenario.
We conclude in Sec.~\ref{sec:con}.

\section{Vector dark matter with kinetic function} \label{sec:vec}

\subsection{Inflaton and vector dynamics}

We consider a following model. The inflaton $\phi$ is coupled to the vector boson (or dark photon) $A_\mu$ through the kinetic function $f(\phi)$:
\begin{align}
	\mathcal L = -\frac{f^2(\phi)}{4} F_{\mu\nu} F^{\mu\nu} - \frac{1}{2}m_A^2 A_\mu A^\mu - \frac{1}{2}\partial_\mu \phi \partial^\mu\phi - V(\phi) + \mathcal L_\chi,
	\label{Lagrangian}
\end{align}
where $\mathcal L_\chi$ denotes the Lagrangian for the curvaton, which is decoupled from the inflaton and vector boson sector. It will be discussed in Sec.~\ref{sec:curvaton} and is irrelevant for the inflaton/vector dynamics, so we neglect it in this section.
We assume $\phi$ is not responsible for the generation of the observed large-scale curvature perturbation. We consider a simple quadratic model for concreteness, 
\begin{align}
	V(\phi) = \frac{1}{2}m_\phi^2\phi^2,
        \label{Vphi}
\end{align}
although it is easily extended to more general inflaton potential.
The kinetic function is chosen to be~\cite{Martin:2007ue}
\begin{align}
	f(\phi)=\exp\left(-\frac{\gamma}{8}\frac{\phi^2}{M_{\rm Pl}^2}\right),
	\label{kinetic}
\end{align}
with some constant $\gamma$, so that the kinetic function scales as some powers of the cosmic scale factor $a(t)$ as the inflaton rolls down the potential.\footnote{
    For the inflaton potential $V(\phi)\propto \phi^n$, we can choose $f(\phi)=\exp\left(-\frac{\gamma}{4n}\frac{\phi^2}{M_{\rm Pl}^2}\right)$ to obtain the same time dependence of $f(\phi)$. See Ref.~\cite{Martin:2007ue} for more general inflaton potential.
}
The equations of motion of the inflaton and the vector, in terms of the ``physical'' field $\overline A_i \equiv f A_i/a$, are given by
\begin{align}
	&\ddot\phi + 3H\dot\phi + \partial_\phi V \left(1 + \frac{\gamma R_A}{2\epsilon_V}\right)=0, \label{phi_eom}\\
	&\ddot{{\overline A_i}}+ 3H \dot{{\overline A_i}} + \left(\frac{m_A^2}{f^2} - \frac{(\alpha+4)(\alpha-2)}{4}H^2 + \frac{2-\alpha}{2}\dot H \right)\overline A_i = 0,
\end{align}
where we have assumed $f^2 \propto a^\alpha$, $R_A \equiv \rho_A/\rho_\phi$, $H=\dot a/a$ is the Hubble parameter and $\epsilon_V\equiv \frac{M_{\rm Pl}^2}{2}\left(\frac{\partial_\phi V}{V}\right)^2 = \frac{2M_{\rm Pl}^2}{\phi^2}$. The vector energy density is given as
\begin{align}
	\rho_A = \frac{f^2 \dot A_i^2}{2a^2} + \frac{m_A^2 A_i^2}{2a^2} \simeq \frac{1}{2}\left[ (1+\alpha)^2 H^2 \overline{A_i}^2 + \frac{m_A^2 \overline{A_i}^2}{f^2} \right],
\end{align}
where in the last similarity we assumed $f\propto a^{\alpha/2}$ and $H\sim {\rm const.}$ $(\gg m_A)$ during inflation.\footnote{
	It is convenient to note that $af^2 \dot A_i \sim {\rm const.}$ for $m_A\ll H$.
}
In the standard slow-roll inflation we have $\alpha=\gamma$. Thus we see that the Hubble mass term for $\overline A_i$ is non-positive for $\gamma\leq -4$ and we consider the case $\gamma < -4$ and $|\gamma+4|\ll 1$ separately below, although it will be turned out that the case of $\gamma<-4$ will produce too large statistical anisotropy and it is not appropriate for phenomenological purpose.\footnote{
	Another possible option is $\gamma \geq 2$. In this case we should also introduce a mass function $ h(\phi) $, so that the vector mass term is replaced as $m_A^2 \to h^2(\phi) m_A^2$ in order to obtain the consistent vector DM abundance, while the isocurvature perturbation of the longitudinal mode is greatly enhanced~\cite{Nakayama:2020rka}. To avoid such a further complexity, we do not consider the case of $\gamma\geq 2$ in this paper.
}

\subsubsection{$\gamma<-4$ : Anisotropic inflation}

For $\gamma < -4$ the vector field obtains effective negative Hubble mass and hence the vector field as well as its energy increases during the slow-roll inflation, unless $\gamma$ is very close to $-4$. Eventually, the backreaction of the vector field to the inflaton becomes important. At this stage, the so-called anisotropic inflation happens~\cite{Watanabe:2009ct,Soda:2012zm,Maleknejad:2012fw}.
In the anisotropic inflation regime, the vector energy density is given by
\begin{align}
	 R_A = -2\epsilon_V \frac{\gamma+4}{\gamma^2}.
	\label{RA_an}
\end{align}
In this regime we have $\alpha=-4$ independently of the value $\gamma$, in contrast to the standard slow-roll inflation regime. 
Note that $\epsilon_V(\tau_{\rm end}) \sim 1$ at the end of inflation and hence $R_A$ becomes close to unity unless $|\gamma|$ is much larger than unity. 
It will produce too large statistical anisotropy for the curvaton-induced curvature perturbation, as we will see in the next section. 


\subsubsection{$|\gamma+4|\ll 1$ : Slow-roll inflation}

On the other hand, if $|\gamma + 4| \ll 1$, the effective (negative) Hubble mass term of the vector field becomes close to zero and it will take longer and longer time to settle into the configuration given in (\ref{RA_an}). We should also take account of the quantum generation of long-wave fluctuations of the vector field during inflation, which are summed up to classical vector background just as in the case of minimal massless scalar~\cite{Bartolo:2012sd}. Practically, therefore, we can take $\overline A_i$ or $\rho_A$ just to be a constant\footnote{
	More precisely, $\overline A \propto a^{-(\gamma+4)/2}$. 
} during inflation keeping the initial condition given by hand for $|\gamma+4| \ll 1$.
In this case, in order for the vector field not to significantly affect the inflaton dynamics, we need
\begin{align}
	R_A\simeq \frac{3\overline A_i^2}{2M_{\rm Pl}^3} \lesssim \frac{\epsilon _V}{2}.
	\label{RA_sl}
\end{align}
In this case we can just take $\alpha=\gamma$. 
For the simple quadratic potential for the inflaton (\ref{Vphi}), $\epsilon_V = 1/(2N)$ with the e-folding number $N \sim 50$--$60$ when the observable scales exit the horizon. Thus Eq.~(\ref{RA_sl}) leads to a constraint like $R_A\lesssim 5\times 10^{-3}$. 

\subsection{Vector boson abundance}

Just after the end of inflation, the vector energy density is given by $\rho_A \simeq R_A \rho_\phi$.
After that, we assume that the inflaton exhibits coherent oscillation around the potential minimum, which behaves as non-relativistic matter, and eventually it decays into radiation at $3H = \Gamma_\phi$, where $\Gamma_\phi$ is the decay rate of the inflaton, as discussed in the next subsection. The reheating temperature $T_{\phi}$ is then given by $T_{\phi}=(10/\pi^2 g_*)^{1/4}\sqrt{\Gamma_\phi M_{\rm Pl}}$.
The vector field scales as follows. For $H > m_A$ one can neglect the vector mass term and we obtain  $\overline A \propto a^{(3w-1)/2}$, where $w$ is the equation of state parameter ($w=0$ during inflaton coherent oscillation and $w=1/3$ during the radiation-dominated era). It leads to $\rho_A \simeq H^2 \overline A^2/2 \propto a^{-4}$.
On the other hand, for $H < m_A$, the coherent oscillation of the vector field starts and we obtain $\overline A \propto a^{-3/2}$ and $\rho_A \sim m_A^2\overline A^2/2 \propto a^{-3}$, as expected from the behavior of the ordinary non-relativistic matter. 
In this oscillating regime, the vector field pressure is isotropic and it does not induce the anisotropic expansion.

\begin{figure}
\begin{center}
   \includegraphics[width=16cm]{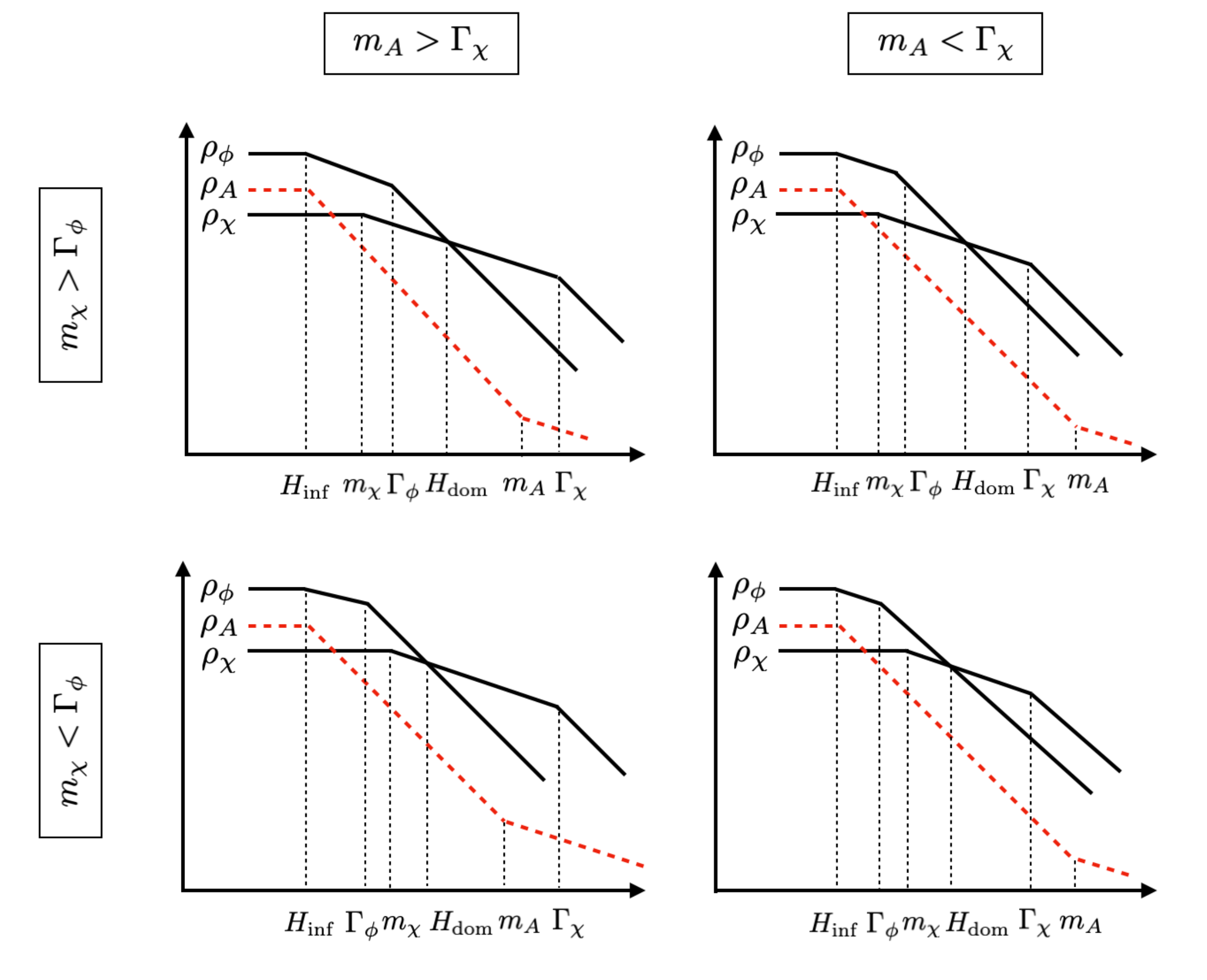}
  \end{center}
  \caption{Schematic view of the time evolution of energy density of the inflaton $\rho_\phi$, curvaton $\rho_\chi$ and vector boson $\rho_A$. Hubble parameters at several characteristic moments are also shown. Note that $\rho_A \propto a^{-4}$ for $H>m_A$ and $\rho_A \propto a^{-3}$ for $H<m_A$.}
  \label{fig:time}
\end{figure}

As already explained in the Introduction and will be discussed in the next section in detail, we introduce a curvaton field $\chi$ to explain the observed density perturbation. The curvaton dominates the universe before it decays at $3H=\Gamma_\chi$ with $\Gamma_\chi$ being the curvaton decay rate,\footnote{
	To be precise, what is needed is the curvaton energy density must be larger than about 10 percent of the total energy density when the curvaton decays in order to avoid too large non-Gaussianity. See next section for more detail. 
} and hence it dilutes the vector boson abundance due to the extra entropy production.
Suppose that the initial curvaton field value is $\chi_i$ and it begins to oscillate at $H=m_\chi$, with $m_\chi$ being the curvaton mass, the curvaton dominates the universe at $H=H_{\rm dom}$, where
\begin{align}
	H_{\rm dom}= {\rm min}\left[\Gamma_\phi, m_\chi\right]\times \left(\frac{\chi_i}{\sqrt{6}M_{\rm Pl}}\right)^4.
	\label{Hdom}
\end{align}
In order for the curvaton to explain the observed density perturbation of the universe, we need $H_{\rm inf}/(\pi \chi_i) \sim 5\times 10^{-5}$ where $H_{\rm inf}$ denotes the Hubble scale during inflation (see Sec.~\ref{sec:curvaton}).\footnote{
    The Hubble scale during inflation when the observable scales exit the horizon, $H_{\rm inf}$, is related to the inflaton mass $m_\phi$ as $H_{\rm inf} \simeq 6m_\phi$ for the quadratic potential (\ref{Vphi}).
} Then $H_{\rm dom}$ is given by
\begin{align}
	H_{\rm dom}\simeq {\rm min}\left[\Gamma_\phi, m_\chi\right]\times \left(\frac{H_{\rm inf}}{9\times 10^{14}\,{\rm GeV}}\right)^4.
\end{align}
See Fig.~\ref{fig:time} for schematic view of the time evolution of energy density of the inflaton $\rho_\phi$, curvaton $\rho_\chi$ and vector boson $\rho_A$.
By definition, $m_{\phi}> (m_\chi,\Gamma_\phi) > H_{\rm dom} \gtrsim \Gamma_\chi$ is satisfied for $\chi_i\lesssim M_{\rm Pl}$.
As will be shown in Sec.~\ref{sec:iso}, we also need $m_A \lesssim H_{\rm dom}$ in order to avoid the too large residual correlated isocurvature perturbation.
Thus the four cases shown in Fig.~\ref{fig:time} are enough to cover all the relevant possibilities.

The present vector coherent energy density, divided by the entropy density $s$, is evaluated as
\begin{align}
	\frac{\rho_A}{s} =\begin{cases}
		\displaystyle R_A^{\rm (end)}\frac{3 T_{\chi}}{4}\left(\frac{\Gamma_\phi}{m_\phi}\right)^{2/3}\left(\frac{m_A}{H_{\rm dom}}\right)^{2/3} & {\rm for}~m_A>\Gamma_\chi \vspace{0.2cm} \\ 
		\displaystyle R_A^{\rm (end)}\frac{3 T_{m_A}}{4}\left(\frac{\Gamma_\phi}{m_\phi}\right)^{2/3} \left(\frac{\Gamma_\chi}{H_{\rm dom}}\right)^{2/3}  & {\rm for}~m_A<\Gamma_\chi
	\end{cases},
 \label{rhoAs}
\end{align}
where $T_{m_A} \equiv (10/\pi^2 g_*)^{1/4}\sqrt{m_A M_{\rm Pl}}$ and $T_\chi\equiv (10/\pi^2 g_*)^{1/4}\sqrt{\Gamma_\chi M_{\rm Pl}}$ is the temperature after the curvaton decay.
This expression does not explicitly depend on whether $\Gamma_\phi$ is larger or smaller than $m_\chi$, which is implicitly contained in $H_{\rm dom}$.
Note that $R_A^{\rm (end)}$ in (\ref{rhoAs}) represents $R_A$ evaluated at the end of inflation. Practically, however, we can just take $R_A$ to be a constant during inflation as far as we consider the case of $|\gamma+4|\ll 1$.

\subsection{Reheating} 

The interaction between the inflaton and vector boson $A_\mu$ in (\ref{Lagrangian}) with the kinetic function (\ref{kinetic}) does not lead to the complete decay of the inflaton, since the coupling is quadratic in $\phi$ and the decay rate vanishes in the limit $\phi\to 0$. 
Still it leads to the nonthermal production of vector boson. The effective decay rate (or it may be interpreted as the annihilation of the inflaton) into the vector boson pair is given by\footnote{
	This is the same order of the purely gravitational production rate during reheating~\cite{Ema:2015dka,Ema:2016hlw,Ema:2019yrd}.
}
\begin{align}
	\Gamma_{\phi\phi\to AA}\sim \frac{1}{16\pi} \frac{\left<\phi^2\right> m_\phi^3}{M_{\rm Pl}^4}.
\end{align}
This nonthermal production is most efficient just after the end of inflation due to the $\left<\phi^2\right>$ dependence. 
The nonthermal vector boson abundance, in terms of the number-to-entropy density ratio, is given by
\begin{align}
	\left(\frac{n_A}{s}\right)_{\rm nonthermal} \sim \frac{3\sqrt{3}}{32\pi}\frac{m_\phi T_{\phi}}{M_{\rm Pl}^2} \Delta
	\simeq 9\times 10^{-19}  \left(\frac{m_\phi}{10^{10}\,{\rm GeV}}\right)
	 \left(\frac{T_{\phi}}{10^{10}\,{\rm GeV}}\right) \Delta,
\end{align}
where $\Delta \leq 1$ denotes the dilution factor due to the curvaton decay.
In order to evaluate the present energy density of nonthermally produced vector bosons, the vector boson mass $m_A$ should be multiplied if it is non-relativistic, or the (kinetic) energy should be multiplied if it is relativistic. In either case, it is negligibly small in the case of light vector boson $m_A \lesssim 1\,{\rm eV}$, which we are interested in.

In order for the inflaton to decay into light degrees of freedom, we need to introduce some additional couplings of the inflaton. 
One of the possible interaction is of the form
\begin{align}
	\mathcal L = -\frac{1}{4}\exp\left(\frac{\phi}{\Lambda}\right) F^{\rm (EM)}_{\mu\nu}F^{\rm (EM) \mu\nu},  \label{phiFF}
\end{align}
where $F^{\rm (EM)}_{\mu\nu}$ denotes the electromagnetic field strength tensor of the Standard Model photon.\footnote{
	The discussion does not change if it is replaced by the Standard Model hypercharge photon.
}
Note that introduction of this term does not amplify the Standard Model photon during inflation and hence does not affect the inflaton dynamics. Note also that large value of $\phi$ during inflation means that the electromagnetic coupling is much weaker than the present universe. 
After the inflation ends, the inflaton oscillates around the potential minimum $\phi=0$. The decay rate into the Standard Model photon is then given by $\Gamma_\phi = m_\phi^3/(64\pi \Lambda^2)$. Thus the reheating temperature after the inflaton decay is estimated as
\begin{align}
	T_{\phi}\simeq 1.4\times 10^4\,{\rm GeV} \left(\frac{m_\phi}{10^{10}\,{\rm GeV}}\right)^{3/2} \left(\frac{M_{\rm Pl}}{\Lambda}\right).
	\label{TR}
\end{align}
Another simple option\footnote{
    One may also consider the inflaton coupling to the Standard Model Higgs boson $H$ as $\mathcal L = \mu\phi|H|^2$. In this case, one must be careful about the destabilization of the Higgs vacuum due to the amplification of the Higgs fluctuations during the preheating~\cite{Herranen:2015ima,Ema:2016kpf,Kohri:2016wof}. In particular, strong upper bound on the trilinear coupling $\mu$ is obtained as $|\mu| \ll m_\phi^2/ M_{\rm Pl}$~\cite{Enqvist:2016mqj,Ema:2017loe}. Still the reheating temperature can be sizable.
} is to introduce a coupling to an extra massless fermion $\psi$ as $\mathcal L = y \phi \bar\psi\psi$. In this case the inflaton decay rate is given by $\Gamma_\phi = y^2m_\phi/(8\pi)$. To avoid the radiative correction to the inflaton potential, which would affect the inflaton dynamics, $y\lesssim 10^{-4}$ is required. Still we can have the reheating temperature much higher than (\ref{TR}).\footnote{
    If $\psi$ is massless and does not interact with the Standard Model sector, it acts as dark radiation. It is harmless if the curvaton dominates the universe and decays into the Standard Model sector.
}

Eventually the curvaton $\chi$ may dominate the universe and hence it must decay into the Standard Model radiation. The curvaton $\chi$ can also have a similar coupling to the photon (\ref{phiFF}). 
In order to avoid the radiative correction to the curvaton potential, it may be better to regard the curvaton as axion-like field. Then it may have an axionic coupling to the photon,
\begin{align}
	\mathcal L = -\frac{\chi}{8\Lambda} \epsilon^{\mu\nu\rho\sigma} F^{\rm (EM)}_{\mu\nu} F^{\rm (EM)}_{\rho\sigma}. \label{chiFF}
\end{align}
In either case, the reheating temperature $T_\chi$ after the curvaton decay is expressed as (\ref{TR}) by just replacing $T_{\phi}\to T_\chi$ and $m_\phi \to m_\chi$.
The following discussion are independent of specific mechanisms of the inflaton/curvaton decay.

\section{Constraints}  \label{sec:const}

\subsection{Anisotropic curvature perturbation from curvaton}  \label{sec:curvaton}

The metric in the presence of the homogeneous vector background is given by the following Bianchi-I form:
\begin{align}
	ds^2 = -dt^2 +a^2(t)\left[e^{-4\sigma}dx^2 + e^{2\sigma}(dy^2+dz^2) \right],
\end{align}
where the $x$ axis is taken to be the direction of the vector condensate $\vec A$ without loss of generality. The anisotropic expansion rate, $\Sigma\equiv\dot\sigma$ satisfies the following equations,
\begin{align}
	&H^2 = \Sigma^2 + \frac{1}{3M_{\rm Pl}^2}(\rho_\phi + \rho_A), \label{Fried} \\
	&\dot\Sigma + 3H\Sigma = \frac{1}{3M_{\rm Pl}^2} \left(\frac{f \dot A_x}{a}\right)^2 e^{4\sigma}. \label{dotSigma}
\end{align}
The right-hand side of Eq.~(\ref{dotSigma}) is approximately constant during inflation and $\Sigma$ approaches to
\begin{align}
	\frac{\Sigma}{H} \simeq \frac{2\rho_A}{3\rho_\phi}=\frac{2 R_A}{3}.
	\label{Sigma}
\end{align}
After inflation ends, $f\sim 1$ and the right-hand side of Eq.~(\ref{dotSigma}) quickly decreases as $a^{-4}$ and hence it is safely neglected. Thus $\Sigma \propto a^{-3}$ thereafter, meaning that the universe becomes isotropic.
Its contribution to the Friedmann equation (\ref{Fried}) also becomes negligible quickly. Thus its effect on the Big-Bang nucleosynthesis or CMB is negligible~\cite{Akarsu:2019pwn}.\footnote{
	Note that $\sigma$ approaches to some constant $\sigma_{\rm end}$ in the late universe. The constant value of $\sigma_{\rm end}$ itself does not indicate observable effects since constant $\sigma$ is absorbed by the redefinition of coordinates to reproduce the isotropic universe. However, time evolution of $\sigma$ can have nontrivial effects as discussed below. 
}
Still it may have a significant impact on the property of the curvature perturbation generated by the curvaton. As we will see below, it leads to the statistical anisotropy in the curvature perturbation power spectrum.

Let us discuss the curvature perturbation generated by the curvaton. We are assuming the existence of the curvaton field, $\chi$, which has no direct couplings to the inflaton and vector field:
\begin{align}
	\mathcal L_\chi = -\frac{1}{2}\partial_\mu\chi \partial^\mu\chi - V(\chi).
\end{align}
Defining $\widetilde\chi\equiv a\chi$, the mode expansion is given by~\cite{Birrell:1982ix}
\begin{align}
	\widetilde\chi(t,\vec x) = \int \frac{d^3k}{(2\pi)^3}\left[ a_{\vec k} \widetilde \chi_{\vec k} +  a^\dagger_{-\vec k} \widetilde \chi^*_{\vec k} \right]e^{i\vec k\cdot\vec x},
\end{align}
where the creation/annihilation operator satisfies the commutation relation $\left[a_{\vec k}, a^\dagger_{\vec k'} \right] = (2\pi)^3\delta(\vec k-\vec k')$
and the mode function satisfies the equation of motion as
\begin{align}
	\widetilde\chi_{\vec k}'' + \left( a^2(\tau) p^2(\tau)-\frac{a''}{a} + a^2m_\chi^2\right) \widetilde\chi_{\vec k} = 0,
	~~~~~~p^2(\tau) = a^{-2}(\tau) \left(  e^{4\sigma} k_x^2 + e^{-2\sigma}\vec k_\perp^2 \right),
	\label{chik}
\end{align}
where $\vec k_\perp = (0,k_y,k_z)$, $\vec p$ denotes the physical momentum and the prime denotes the derivative with respect to conformal time $\tau = \int dt/a$.
Eq.~(\ref{chik}) apparently looks isotropic in terms of the physical momentum $\vec p$, but one should note that $\vec p$ evolves differently depending on its direction due to the anisotropic expansion factor, $e^{4\sigma}$ and $e^{-2\sigma}$ in Eq.~(\ref{chik}), since $\sigma$ is time dependent. For example, the modes $\vec k=(k,0,0)$ and $\vec k=(0,0,k)$ will result in different magnitude of the physical wavenumber after inflation, which may be viewed as the statistical anisotropy.

To see this, let us solve Eq.~(\ref{chik}). 
One subtlety is the initial condition. We assume that the universe before the observable scales exit the horizon is nearly isotropic and $N$ is not much larger than $60$. It is possible that the vector condensate $\overline A$ is almost zero initially (and hence the universe is isotropic) and formed thereafter due to the inflaton motion in the last $N$ e-foldings.\footnote{
    Ref.~\cite{Gumrukcuoglu:2007bx} considered the case where there is no source (i.e. the vector condensate in our case) for the anisotropic expansion. In such a case, the universe becomes more and more anisotropic as time goes back, and the treatment of initial condition is more subtle.   
} 
Although we cannot solve the equation (\ref{chik}) in a conventional analytical way with the Hankel function since $ap$ is time dependent, we can derive an approximate solution by assuming that $\chi_{\vec k}$ freezes out when the mode exits the horizon.
Then the solution to Eq.~(\ref{chik}) is approximately given by
\begin{align}
	\chi_{\vec k} \to
	\begin{cases}
	\displaystyle\frac{1}{a\sqrt{2 ap}} e^{-i ap \tau} \simeq  \frac{1}{\sqrt{2k}a} e^{-i k\tau} & {\rm for}~~ap\tau \to -\infty\\
	\displaystyle\frac{i H_{\rm exit}}{\sqrt{2(ap)^3_{\rm exit}}} & {\rm for}~~ap\tau \to 0
	\end{cases},
\end{align}
where the subscript ``exit'' represents that it is evaluated at the horizon exit; $ap = -1/\tau$.
Therefore, by noting $H_{\rm exit}\simeq H_{\rm inf}(ap/k_0)^{\dot H/H^2}$ with $k_0$ being some reference constant, we obtain
\begin{align}
	\left|\chi_{\vec k}\right|^2 \simeq \frac{H_{\rm inf}^2}{2k^3} \left(\frac{k}{k_0}\right)^{n_s-1} \left[1+2(n_s-4)\sigma_k -3(n_s-4)\sigma_k\sin^2\theta_k\right],
	\label{chi2}
\end{align}
around the observable scales, where $\theta_k$ is defined as the angle between the $x$-axis and the momentum vector $\vec k$, $n_s$ is the scalar spectral index given by $n_s\simeq 1+2\dot H/H^2 + 2m_\chi^2/(3H_{\rm inf}^2)$ and it is consistent with the observation for appropriate choice of the inflaton and curvaton potential\footnote{
    For the quadratic inflaton potential (\ref{Vphi}), in order to obtain the red spectral index of the curvature perturbation, the curvaton needs to be tachyonic during inflation, by interpreting $m_\chi^2$ as the effective mass during inflation that may differ from the present-day mass, and more careful treatment is required to calculate the curvature perturbation. It is actually possible to obtain the consistent spectral index with sizable amount of non-Gaussianity~\cite{Kawasaki:2008mc,Chingangbam:2009xi,Kawasaki:2011pd,Kawasaki:2012gg}.
    On the other hand, if the inflaton potential is quartic $V(\phi)\propto \phi^4$, the scalar spectral index may fall into the value consistent with the observation for $m_\chi\ll H_{\rm inf}$. In this case, the calculation of the vector DM abundance remains the same if the quadratic inflaton mass term comes to dominate at the end of inflation.
} 
and $\sigma_k$ means its value at the horizon crossing of the mode $\vec k$. 
The statistically anisotropic contribution, represented by a term proportional to $\sin^2\theta_k$ in Eq.~(\ref{chi2}), appears because the timing of horizon exit is different depending on the direction of the momentum $\vec k$ due to the anisotropic expansion.

Note that we want to know the power spectrum in terms of the physical wavenumber. After inflation ends, the anisotropic expansion stops and hence the physical wavenumber thereafter is just its value at the end of inflation, $\vec p_{\rm end}$, scaled by the overall scale factor $a$. 
Thus we express the power spectrum in terms of $\vec k_{\rm end}\equiv a_{\rm end}\vec p_{\rm end}$, which is given by
\begin{align}
	\vec k_{\rm end} = a_{\rm end}\vec p_{\rm end} = (e^{2\sigma_{\rm end}}k_x,~e^{-\sigma_{\rm end}}\vec k_\perp).
\end{align}
We then define the power spectrum as
\begin{align}
	\left<\chi^2\right>=\int\frac{d^3k}{(2\pi)^3}\left|\chi_{\vec k}\right|^2 
	= \int d\ln k_{\rm end} \frac{d\cos\theta_k}{2} \mathcal P_\chi(\vec k_{\rm end}),
\end{align}
where
\begin{align}
	\mathcal P_\chi(\vec k_{\rm end}) \simeq \left(\frac{H_{\rm inf}}{2\pi}\right)^2 \left(\frac{k_{\rm end}}{k_0}\right)^{n_s-1}
	\left\{1+\left[3(n_s-1)(\sigma_{\rm end}-\sigma_{k})+9\sigma_k\right]\sin^2\theta_k\right\}.
\end{align}
The extra contribution to the statistical anisotropy arises due to the breaking of scale invariance. As explained above, the evolution of the wavevector $\vec k$ after the horizon exit is different depending on its angle to the $x$-axis. Therefore, if the fluctuation amplitude depends on the time of horizon exit (i.e., if the scale invariance is broken), it will result in the nontrivial angular dependence in terms of the present physical wavenumber. 
Note that we neglected the time dependence of $\theta_k$ since it will only lead to higher order corrections in $\sigma$.
Assuming that the curvaton dominates the energy density of the Universe when it decays into radiation,\footnote{
	If the curvaton energy density is subdominant at the decay, it yields large non-Gaussianity~\cite{Lyth:2001nq,Lyth:2002my}. To avoid too large non-Gaussianity, the curvaton energy fraction should be larger than $\sim 0.1$ when it decays. See, however, Refs.~\cite{Byrnes:2011gh,Mukaida:2014wma} for models to avoid this restriction.
} the curvature perturbation is given by
\begin{align}
	\mathcal P_\zeta^{\rm (curv)}(\vec k_{\rm end}) \simeq \mathcal P_{\zeta 0}^{\rm (curv)}(k_{\rm end})\left[ 1+g^{\rm (curv)}\sin^2\theta_k \right],
	\label{P_curv}
\end{align}
where the isotropic part $P_{\zeta 0}^{\rm (curv)}$ and the statistical anisotropy parameter $g^{\rm (curv)}$ are given by
\begin{align}
 & \mathcal P_{\zeta 0}^{\rm (curv)}(k_{\rm end}) = \left(\frac{H_{\rm inf}}{\pi \chi_i}\right)^2 \left(\frac{k_{\rm end}}{k_0}\right)^{n_s-1},\\
 & g^{\rm (curv)} = 3(n_s-1)(\sigma_{\rm end}-\sigma_{k})+9\sigma_k,
 \label{g_curv}
\end{align}
with $\chi_i$ being the initial field value of the curvaton. It should explain the observed value, $\mathcal P_\zeta^{\rm (curv)}(k_0)=\mathcal P_\zeta^{\rm (obs)}(k_0)\simeq 2.1\times 10^{-9}$, while the statistical anisotropy parameter should be smaller than about $0.01$ from CMB observations~\cite{Planck:2018jri}. 
While $|\sigma_{\rm end}-\sigma_k|\sim R_A N$ with $N \gtrsim 60$ being the e-folding number after the horizon exit of the observable scale, $\sigma_k$ itself can be much larger if the anisotropic expansion lasts long enough before the last $60$ e-foldings. Here we just assume that the anisotropic expansion phase roughly coincides with the last $60$ e-foldings.\footnote{
    It is indeed possible that the inflaton potential or the kinetic function slightly deviates from the simple form (\ref{Vphi}) or (\ref{kinetic}), which would give the negative effective Hubble mass to the vector boson so that the vector condensate develops and eventually it approaches to the simple form (\ref{Vphi}) or (\ref{kinetic}) and $R_A={\rm const.}$ thereafter.
}
In such a case we have vanishingly small $\sigma_k$ at the present cosmological scales and we can approximate $g^{\rm (curv)} \simeq 3(n_s-1)(\sigma_{\rm end}-\sigma_{k})$. Taking account of the observed value $n_s-1\sim 0.035$~\cite{Planck:2018jri}, we estimate the constraint as $0.1\times \sigma_{\rm end} \sim 0.1 R_A N \lesssim 0.01$, which reads $R_A \lesssim 10^{-3}$. 

It may be worth mentioning that $\sigma_k$ becomes larger and the second term in (\ref{g_curv}) becomes dominant 
for smaller scales. One should also note that the constraint on $R_A$ can be much stronger if the anisotropic expansion phase with constant $R_A$ would be longer before the observable scales exit the horizon.

\subsection{Anisotropic curvature perturbation from inflaton} 

In our scenario, the curvaton is the main source of the observed curvature perturbation, but still the inflaton $\phi$ contributes to the curvatrue perturbation, which is statistically anisotropic due to the relatively strong interaction with the vector boson. Let us derive a condition that this contribution does not exceed  the curvaton contribution.
Below we focus on the case of $|\gamma+4|\ll 1$ as explained in Sec.~\ref{sec:curvaton}.

It is convenient to make use of the $\delta N$-formalism for estimating the curvature perturbation~\cite{Sasaki:1995aw,Wands:2000dp,Lyth:2004gb}, which was extended to the case of anisotropic inflation in Ref.~\cite{Abolhasani:2013zya}.
Let us rewrite the equation of motion of the inflaton (\ref{phi_eom}) in the slow-roll limit in terms of the e-folding number $N$ as
\begin{align}
	\frac{d \phi^2}{dN} = 4M_{\rm Pl}^2 \left(1+ \frac{\gamma \rho_A}{2m_\phi^2 M_{\rm Pl}^2} \right).
\end{align}
Since $\rho_A$ changes very slowly ($\rho_A \propto a^{-(\gamma+4)}$), we can integrate this equation as
\begin{align}
	\phi^2(N) - \phi_{\rm end}^2 \simeq 4M_{\rm Pl}^2 N \left(1+ \frac{\gamma \rho_A}{2m_\phi^2 M_{\rm Pl}^2} \right).
\end{align}
This relation shows how the background motion of the inflaton $\phi$ and the vector $\bar A_i$ affects the e-folding number $N$. What we want to know is how the e-folding number $N$ until the end of inflation\footnote{
	The instance of the end of inflation is naturally defined in the uniform density time slice.
} changes when we perturb the inflaton and vector field amplitude initially. It is read off from this equation as\footnote{
    This equation is easily extended to the case of $V(\phi)\propto \phi^n$. As far as $NR_A \ll 1$, which is actually required from the discussion in the previous subsection, the following estimation is applicable by interpreting $m_\phi^2$ as the effective mass $|\partial_\phi^2 V|$ during inflation.  
}
\begin{align}
	\delta N = \frac{\phi \delta \phi}{2M_{\rm Pl}^2} + N  \frac{18 H_{\rm inf}^2 \overline A_i \delta\overline A_i}{m_\phi^2 M_{\rm Pl}^2}.
\end{align}
The $\delta N$-formalism states that this is equivalent to the curvature perturbation of the universe: $\zeta = \delta N$. 
The power spectrum of the curvature perturbation is obtained by using the following results for the inflaton and vector power spectrum: 
\begin{align}
	&\left< \delta \phi(\vec k) \delta \phi(\vec k') \right> =  \frac{2\pi^2}{k^3} \mathcal P_\phi (k) (2\pi)^3\delta(\vec k+ \vec k'),\\
	&\left< \delta \overline A_i(\vec k) \delta \overline A_j(\vec k') \right> = \frac{2\pi^2}{k^3}\left[\mathcal P_T(\delta_{ij}-\hat k_i \hat k_j) + \mathcal P_L \hat k_i \hat k_j \right] (2\pi)^3\delta(\vec k+ \vec k'),  \label{AA}
\end{align}
where $\mathcal P_\phi (k) \simeq \mathcal P_T (k)  \simeq H^2_{\rm inf}/(2\pi)^2$ while the longitudinal fluctuation is suppressed $ \mathcal P_L (k) \ll \mathcal P_T (k)$ in observable scales~\cite{Nakayama:2019rhg,Nakayama:2020rka}.\footnote{
    Here, we neglect the effect from the anisotropy of the background metric because it gives only a subdominant correction.
}
As a result, the (sub-dominant) curvature perturbation generated by the inflaton is given as
\begin{align}
	\mathcal P_\zeta^{\rm (inf)}(\vec k) =\mathcal P_{\zeta 0}^{\rm (inf)}(k) \left[ 1 + g^{\rm (inf)} \sin^2 \theta_k\right],
\end{align}
where $\theta_k$ is the angle between the wave vector $\vec k$ and the background vector field $\vec A$, and
\begin{align}
	\mathcal P_{\zeta 0}^{\rm (inf)}(k) =\frac{N^2(k)}{6\pi^2} \frac{m_\phi^2}{M_{\rm Pl}^2}.
\end{align}
The statistical anisotropy parameter is given by~\cite{Bartolo:2012sd,Abolhasani:2013zya}
\begin{align}
	g^{\rm (inf)}(k) \simeq \frac{48 N^2(k) R_A(k) }{\epsilon_V(k)},
\end{align}
where the $\tau$ dependence of $\epsilon$ and $R_A$ are converted to the $k$ dependence through the relation $-k\tau=1$, i.e., the mode that exits the horizon at $\tau$.
Note that $N= 50$--$60$ for the present cosmological scales and hence $|g^{\rm (inf)}|$ can be a large number unless $R_A$ is very small. 
We demand that this anisotropic contribution, $\mathcal P_{\zeta 0}^{\rm (inf)} g^{\rm (inf)}$, is smaller than about 1\,\% of the observed curvature perturbation $\mathcal P_\zeta^{\rm (obs)} \simeq 2.1\times 10^{-9}$ from the CMB measurement by the Planck satellite~\cite{Planck:2018jri}. It leads to a constraint
\begin{align}
	m_\phi\lesssim 3\times 10^{10}\,{\rm GeV}\left(\frac{10^{-4}}{R_A}\right)^{1/2},~~~~~~
	H_{\rm inf}\lesssim 2\times 10^{11}\,{\rm GeV}\left(\frac{10^{-4}}{R_A}\right)^{1/2}.
 \label{constraint_ani}
\end{align}
Note that this constraint is much weaker than the case studied in Ref.~\cite{Nakayama:2020rka}, where the inflaton was assumed to generate the observed density perturbation and hence $\epsilon_V$ must be smaller for smaller inflation scale $H_{\rm inf}$.
In our present case, $\epsilon_V$ and $H_{\rm inf}$ are not directly related.

\subsection{Dark matter isocurvature perturbation} \label{sec:iso}

In our model, the vector boson has large scale isocurvature perturbations that are severely constrained by the CMB observations.
As shown in Appendix, the DM isocurvature perturbation in our model is given by
\begin{align}
	S(\vec x) = \frac{\delta\rho_A(\vec x)}{\overline\rho_A} + 3\left( \frac{R_\chi^{\rm (osc)}}{R_\chi^{\rm (dec)}} -1\right) \zeta(\vec x),
	\label{SDM}
\end{align}
where
\begin{align}
	R_\chi = \frac{3\rho_\chi}{4\rho_\phi + 3\rho_\chi},
 \label{Rchi}
\end{align}
with $\rho_\chi$ being the curvaton energy density and $\rho_\phi$ meaning the energy density of the radiation generated by the inflaton decay. The superscript (osc) and (dec) indicate that it should be evaluated when the vector boson begins to oscillate ($3H=m_A$) and when the curvaton decays ($3H=\Gamma_\chi$), respectively.
The first term of (\ref{SDM}) represents the ``intrinsic'' isocurvature perturbation that directly arises from the quantum fluctuation of the vector boson during inflation as given in Eq.~(\ref{AA}). It is uncorrelated with the curvature perturbation $\zeta$. 
The second term of  (\ref{SDM}) represents the ``residual'' isocurvature perturbation that often arises in the curvaton scenario~\cite{Lyth:2002my,Lyth:2003ip,Kitajima:2017fiy}. 
To understand this, let us consider the case where there is no intrinsic fluctuation of DM. In this case, the DM spatial distribution on superhorizon scale is dynamically aligned to the dominant component of the universe at $3H=m_A$, 
since the onset of DM oscillation (i.e. the misalignment DM production) in each patch of the Universe is determined by the Hubble parameter. In other words, the uniform density slice coincides with the uniform DM density slice.
Therefore, if the universe was 
dominated by the inflaton-induced radiation 
at $3H=m_A$, 
the spatial fluctuation of DM follows the one originated from the inflaton 
thereafter, which is clearly the isocurvature mode. It is fully (anti-)correlated with the curvature perturbation.
They both give stringent constraint on our scenario.\footnote{
    (Correlated) baryonic isocurvature perturbation is also strongly constrained. If the baryon asymmetry is created after the curvaton decay, baryons do not have isocurvature perturbations.
} Below we discuss them separately.

\subsubsection{Intrinsic isocurvature perturbation}

First let us consider the intrinsic isocurvature perturbation.
By noting that $\delta\rho_A(\vec x)/\overline\rho_A = 2\delta \overline A_i/\overline A_i$ and using Eq.~(\ref{AA}), we find that the nearly scale invariant isocurvature perturbation power spectrum as~\cite{Nakayama:2019rhg}
\begin{align}	
 \mathcal P^{\rm (int)}_S(\vec k) = \mathcal P^{\rm (int)}_{S0}(k) \sin^2\theta_k,  \label{PS}
\end{align}
where 
\begin{align}	
 \mathcal P^{\rm (int)}_{S0}(k) \simeq \left(\frac{H_{\rm inf}}{\pi \overline A_i}\right)^2 \simeq \frac{3}{2R_A}\left(\frac{H_{\rm inf}}{\pi M_{\rm Pl}}\right)^2.
\end{align}
It should be smaller than $\sim 10^{-2} \mathcal P_\zeta^{\rm (obs)}$ from the CMB measurement by the Planck satellite~\cite{Planck:2018jri}.\footnote{
    There is no availble data for constraint on the statistically anisotropic DM isocurvature perturbation like (\ref{PS}), but we expect that there is no orders-of-magnitude difference between the case of isotropic and anisotropic perturbations.
} It leads to
\begin{align}
	H_{\rm inf} \lesssim 3\times 10^{11}\,{\rm GeV} \left(\frac{R_A}{10^{-4}}\right)^{1/2}.
 \label{constraint_iso}
\end{align}
This is complementary to the constraint (\ref{constraint_ani}) in a sense that the isocurvature constraint becomes severer for small $R_A$ while the constraint (\ref{constraint_ani}) becomes weaker.
Combining these two constraints, we obtain general upper bound as $H_{\rm inf}\lesssim 2\times 10^{11}\,{\rm GeV}$.

\subsubsection{Residual isocurvature perturbation}

Next let us consider the residual isocurvature perturbation which arises if the vector boson starts to oscillate before the curvaton decay.
The factor $R_\chi^{\rm (osc)}/R_\chi^{\rm (dec)} $ is close to unity if the curvaton already dominates the universe when the vector boson starts to oscillate at $3H=m_A$. If, on the other hand, the vector boson begins to oscillate well before the curvaton domination, we have $R_\chi^{\rm (osc)}/R_\chi^{\rm (dec)} \ll 1$, which implies $S/\zeta \simeq -3$. It is clearly too large. 
From the CMB anisotropy measurement one needs~\cite{Planck:2018jri}
\begin{align}
    3\left| \frac{R_\chi^{\rm (osc)}}{R_\chi^{\rm (dec)}} -1\right| \lesssim 0.1.
    \label{constraint_residual}
\end{align}
It indicates that the vector boson must not begin to oscillate much before the curvaton domination: $m_A \lesssim H_{\rm dom}$.
Note again that in order to avoid the too large local non-Gaussianity, we must have $R_\chi^{\rm (dec)} \gtrsim 0.16$.

\subsection{Combined constraints} 


Now let us summarize all the constraints discussed so far and show the consistent parameter regions for the vector coherent oscillation DM. We have many parameters: the initial vector boson density parameter $R_A$, the vector boson mass $m_A$, the inflationary Hubble scale $H_{\rm inf}$ (which is related to the inflaton mass as $H_{\rm inf}\simeq 6m_\phi$), the inflaton decay rate $\Gamma_\phi$ (which is also rewritten in terms of the reheating temperature after inflaton decay $T_{\phi}$), the curvaton initial amplitude $\chi_i$, the curvaton mass $m_\chi$, the curvaton decay rate $\Gamma_\chi$ (which is also rewritten in terms of the reheating temperature after curvaton decay $T_{\chi}$).
They are constrained in nontrivial ways.

\begin{itemize}
\item The vector boson abundance is given by Eq.~(\ref{rhoAs}). It should be consistent with the observed value, $\rho_{\rm DM}/s \simeq 4\times 10^{-10}\,{\rm GeV}$.
\item The backreaction of the vector boson to the inflaton dynamics should not be significant. It requires the condition (\ref{RA_sl}), or $R_A \lesssim 5\times 10^{-3}$. It is comparable or weaker than the constraint from the statistical anisotoropy explained below.
\item The curvaton must explain the observed density perturbation (see Eq.~(\ref{P_curv})). It requires $H_{\rm inf}/(\pi \chi_i) \simeq 5\times 10^{-5}$.
\item The statistical anisotropy of the curvature perturbation must be small enough. It requires $R_A \lesssim 10^{-3}$. See Eq.~(\ref{P_curv}) and texts below it.
\item Non-Gaussianity should be small enough. It requires $R_\chi^{\rm (dec)} \gtrsim 0.16$, which is roughly equal to the condition $\Gamma_\chi \lesssim H_{\rm dom}$, where $H_{\rm dom}$ is given by Eq.~(\ref{Hdom}).
\item The curvaton must decay before the Big-Bang Nucleosynthesis begins. It requires $T_\chi \gtrsim 5\,{\rm MeV}$ or $\Gamma_\chi \gtrsim 10^{-23}\,{\rm GeV}$~\cite{Kawasaki:2000en,Hannestad:2004px,Ichikawa:2005vw,Hasegawa:2019jsa}.
\item The statistically anisotropic inflaton contribution to the total curvature perturbation must be small enough. It imposes a constraint as (\ref{constraint_ani}).
\item The intrinsic isocurvature perturbation must be small enough, which leads to the constraint (\ref{constraint_iso}).
\item The residual isocurvature perturbation must be small enough, which leads to the constraint (\ref{constraint_residual}). It is roughly rephrased as $m_A \lesssim H_{\rm dom}$.
\item Some parameter consistency: $H_{\rm inf} > \Gamma_\phi, m_\chi$.
\end{itemize}

\begin{figure}
\begin{center}
   \includegraphics[width=8cm]{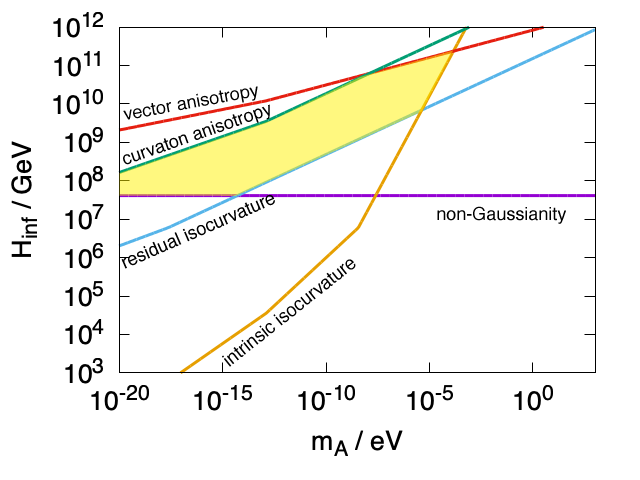}
   \includegraphics[width=8cm]{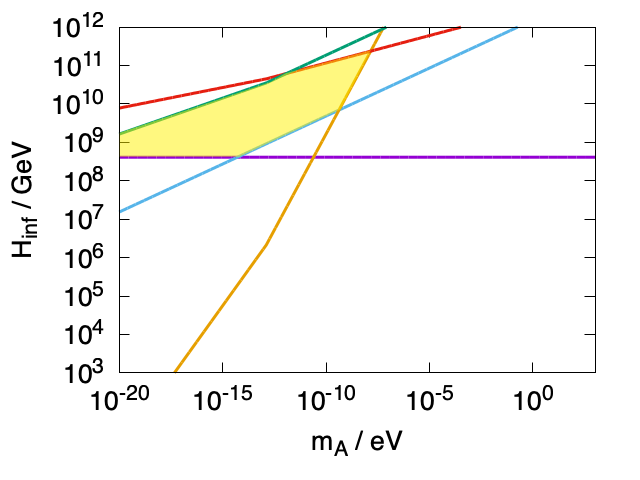}
   \includegraphics[width=8cm]{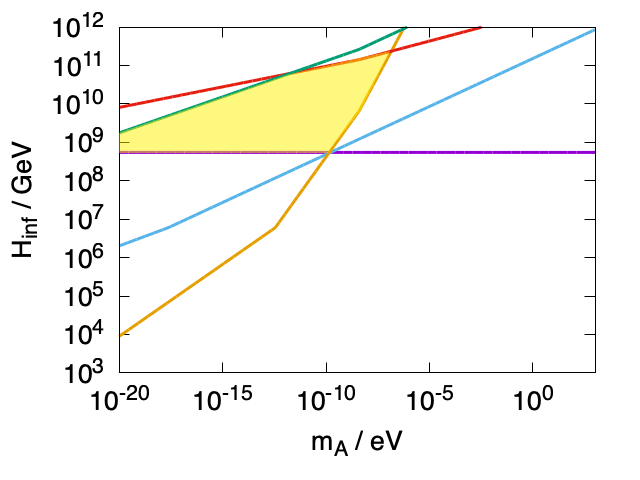}
   \includegraphics[width=8cm]{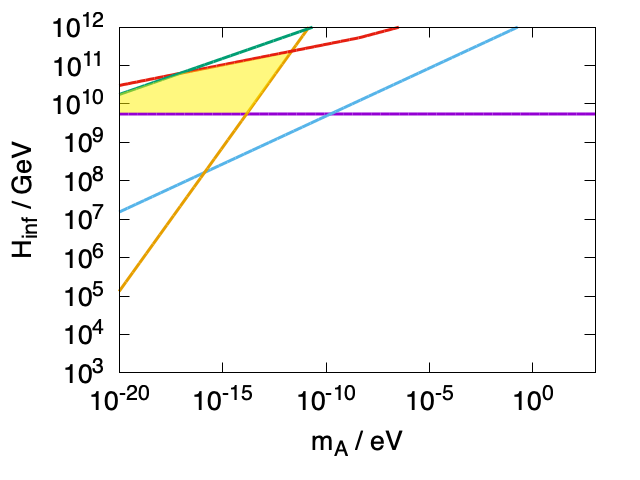}
  \end{center}
  \caption{Allowed parameter region consistent with the vector coherent DM scenario (yellow-shaded) in $H_{\rm inf}$-$m_A$ plane for $m_\chi=10^{6}$ GeV (left panels), $10^2$ GeV (right panels), $T_\chi = 10$ MeV (top panels), 1 GeV (bottom panels). Instant reheating ($\Gamma_\phi=m_\phi$) is assumed.}
  \label{fig:contour}
\end{figure}

\begin{figure}
\begin{center}
   \includegraphics[width=8cm]{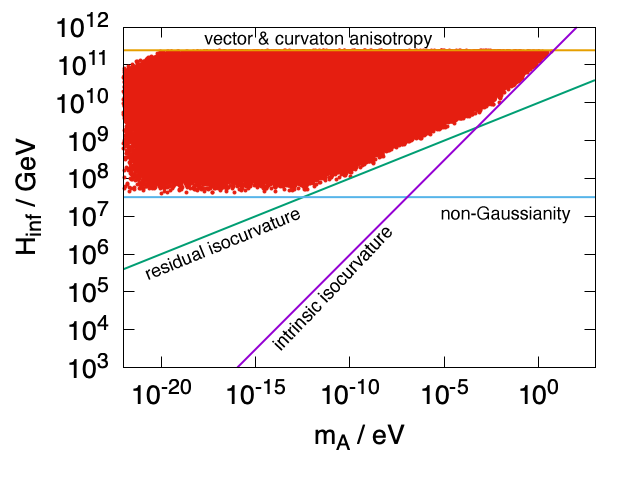}
   \includegraphics[width=8cm]{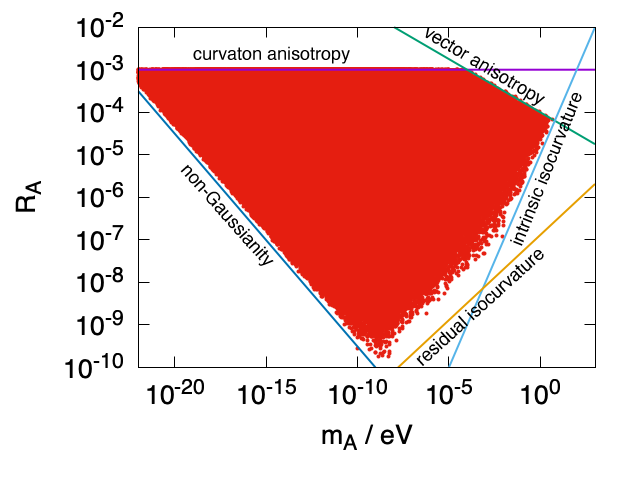}
   \includegraphics[width=8cm]{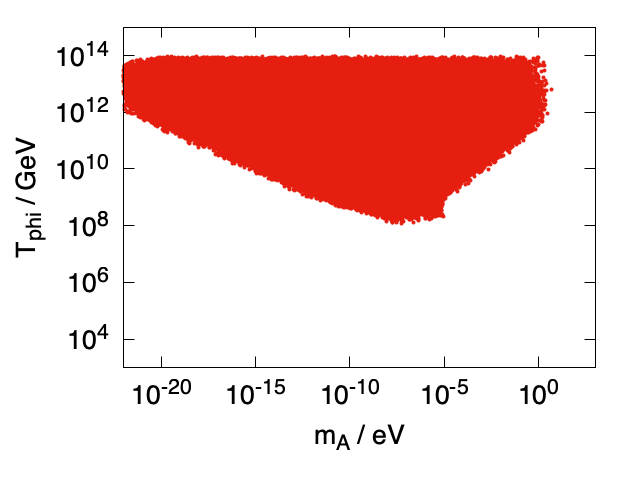}
   \includegraphics[width=8cm]{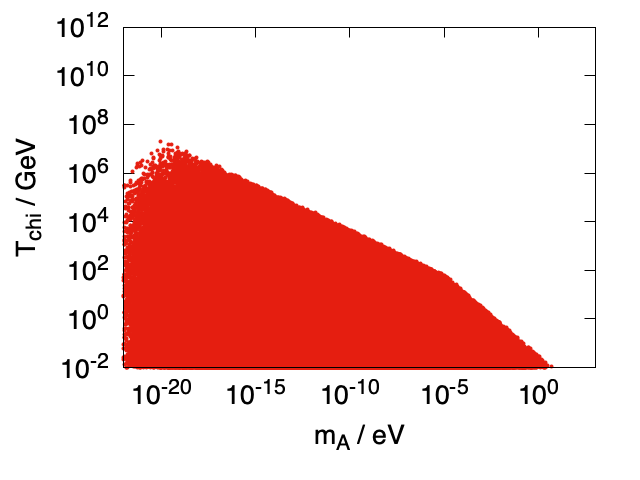}
  \end{center}
  \caption{Scatter plot of allowed parameter region that gives observed DM abundance and curvature perturbation.}
  \label{fig:scatter}
\end{figure}

Figure \ref{fig:contour} shows the allowed region consistent with vector coherent DM scenario in $H_{\rm inf}$-$m_A$ plane (yellow shaded region) which gives observed DM relic abundance and curvature perturbation assuming the instant reheating ($\Gamma_\phi = m_\phi$) with fixed $m_\chi$ and $T_\chi$.
Figure \ref{fig:scatter} shows the scatter plots of allowed parameter region without fixing any parameters.
There are parameter regions consistent with all the constraints mentioned above for the vector boson mass $m_A\sim 10^{-21}$--$1\,{\rm eV}$.
There are lower bounds on the inflationary Hubble scale $H_{\rm inf}$, because low $H_{\rm inf}$ requires low $\chi_i$ for reproducing the observed curvature perturbation and it becomes more difficult for the curvaton to dominate the universe to satisfy the non-Gaussianity bound.
Note that the isocurvature constraint also gives lower bound on $H_{\rm inf}$ due to the nontrivial dependence of $R_A$ on $H_{\rm inf}$ for fixed vector DM abundance.

\section{Conclusions and discussion} \label{sec:con}

We constructed an explicit model for the vector coherent oscillation DM. A basic idea is the introduction of a specific kinetic coupling of the inflaton $\phi$ to the dark photon, based on earlier proposals~\cite{Nakayama:2019rhg}. 
An obstacle to this idea was that the DM isocurvature perturbation and/or the statistical anisotropy of the inflaton curvature perturbation are too large to be consistent with observations~\cite{Nakayama:2020rka}.
Thus we extended the setup to add an extra field, curvaton $\chi$, in order to explain the observed large scale curvature perturbation, while the inflaton is not responsible for it.
Still we find that there are nontrivial constraints from the properties of cosmological fluctuation: intrinsic and residual isocurvature perturbations and also statistical anisotropy.  
We studied in detail this scenario and found that it is indeed possible to realize vector coherent oscillation DM for the very wide mass range $m_A=10^{-21}\,{\rm eV}$--$1\,{\rm eV}$.
In particular, we found a consistent ultra-light vector DM scenario as light as $m_A \sim 10^{-20}\,{\rm eV}$. So far only a few models are known for such an ultra-light vector DM scenario. One is the tachyonic production from the axion coupling. It requires a large axion-dark photon coupling, which needs to be justified with a careful model building~\cite{Agrawal:2018vin}. 
Another one is the production from cosmic strings~\cite{Long:2019lwl}, although it is likely to generate too much gravitational waves exceeding the current pulsar timing constraint~\cite{Kitajima:2022lre}.
Our model is also subject to severe cosmological constraints and tuning of the kinetic function $f(\phi)$, but still there are consistent parameter regions.
Our model provides one concrete example of the consistent vector coherent oscillation DM model and further modification or generalization may be possible. For example, the scalar $\phi$ which enters in the vector kinetic function may need not be the inflaton. 
Such model buildings will be discussed elsewhere.

Finally we comment on possible observable signatures of our scenario. 
As seen from Figs.~\ref{fig:contour} and \ref{fig:scatter}, viable parameter regions are surrounded by several constraints from cosmological observations. In other words, it may be probable that the actual parameter is close to one of these boundaries. Thus at least one of the followings may be observable in future: (statistically anisotropic) isocurvature perturbation, statistically anisotropic curvature perturbation, non-Gaussianity from the curvaton. 
In particular, statistically anisotropic isocurvature perturbation with nearly scale invariant power spectrum is a unique feature of the present model, which would be a smoking-gun signature.
Also there are many ideas and proposals to search for ultra-light vector DM if the vector has a (small) kinetic mixing with the Standard Model photon or if it is the $B-L$ gauge boson with small gauge couplings~\cite{Chaudhuri:2014dla,Graham:2015ifn,Berge:2017ovy,Manley:2020mjq,Michimura:2020vxn,Abe_2021,Chigusa:2023hms}.

Let us also comment on gravitational wave signatures. In our model, statistically anisotropic tensor mode is also predicted in a similar manner to the scalar perturbation discussed in Sec.~\ref{sec:curvaton}. The tensor power spectrum may be expressed as
\begin{align}
    &\mathcal{P}_T(k_{\rm end}) = \mathcal{P}_{T0}(k_{\rm end})(1+g^{(T)} \sin^2\theta_k),\\
    &g^{(T)}=3n_T(\sigma_{\rm end}-\sigma_k)-9\sigma_k,
\end{align}
with $\mathcal{P}_{T0}\simeq 2 H_{\rm inf}^2/(\pi M_{\rm Pl})^2$ and $n_T = -2\epsilon_V$ being the tensor spectral index. It is also a characteristic property of our model, although the inflation scale is severely bounded above and the tensor-to-scalar ratio is too small to be detected in CMB experiments.
There is another production mechanism of the gravitational waves sourced by the vector field~\cite{Fujita:2018zbr}. It is smaller than the vacuum contribution mentioned above by a factor $R_A \sim (\overline A_i/M_{\rm Pl})^2$.

\section*{Acknowledgment}

This work was supported by JSPS KAKENHI Grant Nos. 17H06359 (K.N.), 18K03609 (K.N.), 19H01894 (N.K.), 20H01894 (N.K.), 20H05851 (N.K.), 21H01078 (N.K.), 21KK0050 (N.K.).
This work was supported by World Premier International Research Center Initiative (WPI), MEXT, Japan.

\appendix
\section{Isocurvature perturbation} \label{app:iso}

In this Appendix we make use of the $\delta N$ formalism~\cite{Sasaki:1995aw,Wands:2000dp,Lyth:2004gb} to calculate the DM isocurvature perturbation~\cite{Kawasaki:2008sn,Langlois:2008vk,Kawasaki:2008pa,Kitajima:2017fiy}.
Let $\zeta_i(\vec x)$ be the curvature perturbation on the time slice where the energy density of $i$-th component is uniform, while $\zeta(\vec x)$ be the curvature perturbation on the uniform total density slice. Each $\zeta_i(\vec x)$ is conserved on super-horizon scales as far as its equation of state does not change.
The curvature perturbation $\zeta_i(\vec x)$ coincides with the local e-folding number from the initial spatially flat slice to the final slice where $\rho_i$ is uniform, after subtracting the e-folding number of the background evolution.

First let us consider the uniform density slice at the curvaton $H=\Gamma_\chi$, under the sudden decay approximation. On this slice we have
\begin{align}
	\rho^{\rm (dec)}_{\rm total}  = \rho^{\rm (dec)}_\phi(\vec x) + \rho^{\rm (dec)}_\chi(\vec x)
	= \overline \rho^{\rm (dec)}_\phi e^{4(\zeta_\phi-\zeta)} + \overline \rho^{\rm (dec)}_\chi e^{3(\zeta_\chi-\zeta)},
\end{align}
where $\rho_\phi$ actually means the radiation energy density produced by the inflaton decay. The superscript (dec) reminds us that quantity is evaluated on the slice $H=\Gamma_\chi$ and the overline indicates the unperturbed value. From this we obtain
\begin{align}
	\zeta(\vec x) = R_\chi^{\rm (dec)} \zeta_\chi(\vec x)  + \left(1-R_\chi^{\rm (dec)}\right) \zeta_\phi(\vec x),
	\label{zeta_dec}
\end{align}
where $R_\chi$ is defined in Eq.~(\ref{Rchi}). After the curvaton decay, $\zeta$ is conserved since the universe is just dominated by radiation.

Next let us consider the uniform density slice at $H=m_A$, when the vector boson begins to oscillate. On this slice we have
\begin{align}
	\rho_A^{\rm (osc)}(\vec x) = \overline \rho_A^{\rm (osc)} e^{3(\zeta_A- \zeta^{\rm (osc)} )},
\end{align}
where the superscript (osc) reminds us that quantity is evaluated on the slice $H=m_A$. From this we obtain
\begin{align}
	\zeta_A (\vec x) = \zeta^{\rm (osc)}(\vec x) + \frac{\delta\rho_A^{\rm (osc)}(\vec x)}{3 \overline \rho_A^{\rm (osc)}}.
	\label{zeta_A}
\end{align}
Here $\zeta^{\rm (osc)}(\vec x)$ is given in a similar manner to (\ref{zeta_dec}) as
\begin{align}
	\zeta^{\rm (osc)}(\vec x) = R_\chi^{\rm (osc)} \zeta_\chi(\vec x)  + \left(1-R_\chi^{\rm (osc)}\right) \zeta_\phi(\vec x) .
\end{align}

The DM isocurvature perturbation is defined by
\begin{align}
	S(\vec x) = 3\left(\zeta_A(\vec x)  - \zeta(\vec x) \right).
\end{align}
It is evaluated by substituting Eqs.~(\ref{zeta_dec}) and (\ref{zeta_A}). Since we are interested in the case where the curvaton dominates the total curvature perturbation, we neglect $\zeta_\phi$. Then we obtain
\begin{align}
	S(\vec x) =  \frac{\delta\rho_A^{\rm (osc)}(\vec x)}{\overline \rho_A^{\rm (osc)}} + 3\left( \frac{R_\chi^{\rm (osc)}}{R_\chi^{\rm (dec)}} -1\right) \zeta(\vec x),
\end{align}
which gives Eq.~(\ref{SDM}).

\bibliographystyle{utphys}
\bibliography{ref}

\end{document}